\documentclass[nofootinbib,prd,twocolumn,showpacs,showkeys,preprintnumbers]{revtex4-1}
\usepackage{hyperref,amssymb,amsmath,mathrsfs,bm,graphicx}
\usepackage{multirow}
\usepackage{array}
\usepackage{dblfloatfix}
\usepackage{dblfloatfix}
\usepackage{placeins}
\usepackage{float}
\usepackage[dvipsnames]{xcolor}
\begin{document}

\title {A simple protocol to construct solutions with vanishing complexity by Gravitational Decoupling.}

\author{E. Contreras }
\email{econtreras@usfq.edu.ec}
\affiliation{Departamento de F\'isica, Colegio de Ciencias e Ingenier\'ia, Universidad San Francisco de Quito,  Quito 170901, Ecuador.\\}

\author{Z. Stuchlik}
\email{zdenek.stuchlik@physics.slu.cz}
\affiliation{Research Centre for Theoretical Physics and Astrophysics,
Institute of Physics, Silesian University in Opava, CZ 746 01 Opava, Czech Republic.\\}

\begin{abstract}
In this work we develop a simple protocol to construct interior solutions through Gravitational Decoupling
by the Minimal Gemetric Deformation extended satisfying the vanishing complexity condition. The method is illustrated by using Tolman VII and Tolman IV solutions as isotropic seeds.
\end{abstract}


\maketitle
\section{Introduction}
In recent works \cite{Ovalle:2022yjl,Contreras:2022nji}, the Gravitational Decoupling (GD) ~\cite{Ovalle:2017fgl} thorough the Minimal Geometric Deformation extended (MGD-e) \cite{Ovalle:2019qyi} (see \cite{Casadio:2015jva,Fernandes-Silva:2018abr,Panotopoulos:2018law,daRocha:2019pla,Heras:2019ibr,daRocha:2020rda,daRocha:2020jdj,Tello-Ortiz:2020euy,daRocha:2020gee,Meert:2020sqv,Tello-Ortiz:2021kxg,Maurya:2021huv,Azmat:2021kmv,Maurya:2021zvb,Ovalle:2017wqi,Gabbanelli:2018bhs,Heras:2018cpz,Estrada:2018zbh,Morales:2018urp,Estrada:2018vrl,Ovalle:2018umz,Ovalle:2018ans,Gabbanelli:2019txr,Estrada:2019aeh,Ovalle:2019lbs,Casadio:2019usg,Singh:2019ktp,Maurya:2019noq,Tello-Ortiz:2020ydf,Maurya:2020rny,Rincon:2020izv,Maurya:2020gjw,Zubair:2020lna,Sharif:2020rlt,Ovalle:2020kpd,Ovalle:2020fuo,Estrada:2020ptc,Maurya:2020djz,Meert:2021khi,Maurya:2021aio,Azmat:2021qig,Islam:2021dyk,Afrin:2021imp,Ovalle:2021jzf,Ama-Tul-Mughani:2021ewd,daRocha:2021aww,Maurya:2021qye,Carrasco-Hidalgo:2021dyg,Sultana:2021cvq,daRocha:2021sqd,Maurya:2021yhc,Omwoyo:2021uah,Afrin:2021ggx,Ovalle:2022eqb,Andrade,Dayanandan:2021odo,Contreras:2021yxe} for applications of GD in standard general relativity. For applications in higher dimensions see \cite{Maurya:2022brt,Maurya:2022uqu}, for example) has been proven as an effective route to explore the energy exchange between relativistic fluids supporting a self--gravitating sphere regardless their nature. Indeed, it is well--known that extracting the information on which source dominates over the others is not trivial but, through GD, the accomplishment of such a task is straightforward.

From a more technical point of view, the construction of solutions through GD by MGD-e requires the implementation of two supplementary conditions (such as metric conditions or equations of state) which allow to find the so--called temporal and radial deformation functions. As we shall see later, although the radial deformation can be obtained independently after imposing some suitable equation of state, obtaining the temporal deformation is not straightforward in general. In this work we demonstrate that the complexity factor introduced in \cite{Herrera:2018bww} (for recent developments see \cite{Herrera:2018czt,Sharif:2022wnl,Sharif:2021gsl,Yousaf:2022bwc,Yousaf:2020dkj,Yousaf:2020dci,Zubair:2020poe,Zubair:2020nmx,Maurya:2022cyv,Maurya:2022pef}, for example) can be used to obtain a constraint on the metric functions which allow to obtain the temporal deformation analytically. The interest on the complexity factor has increase in recent years given that it provides useful information about how ``complex'' is a system in comparison with that supported by a homogeneous and isotropic fluid satisfying the so--called vanishing complexity condition. Besides, it provides a non--linear equation of state which allows to complement the required conditions to solve the system of Einstein field  equations.

\par
The paper is organized as follows.
In Section~\ref{sec2}, we review the fundamentals of the GD approach 
to a spherically symmetric system containing two generic sources. In
Section~\ref{sec3}, we introduce the complexity factor and use the vanishing complexity condition to find an analytical expression for the temporal deformation. Besides, we propose a simple protocol to construct well--behaved interior solutions. In sections \ref{tolman} and \ref{tolman4} we test our method by using the well--known Tolman VII and Tolman IV solutions as isotropic seed, respectively. We summarized our results and provide some final comments in the last section.
\section{Gravitational Decoupling}
\label{sec2}
The GD is a well--established formalism that allows either the deconstruction of a complex source or the construction of suitable anisotropic solutions from isotropic seeds. The manner in which it works can be found in plenty of paper in detail so here, we shall summarize the method by following Refs. \cite{Ovalle:2017fgl,Ovalle:2019qyi} . 

Let us consider static and spherically symmetric space--times with a line element parameterized as
\begin{eqnarray}\label{line-element}
ds^{2}=e^{\nu}dt^{2}-e^{\lambda}dr^{2}-r^{2}(d\theta^{2}+\sin^{2}\theta d\phi).
\end{eqnarray}
Using (\ref{line-element}) in Einstein's equations \footnote{We are using units such that $c=G=1$. In this regard, $k^{2}=8\pi$.}
\begin{eqnarray}\label{EFE}
R_{\mu\nu}-\frac{1}{2}g_{\mu\nu}R=k^{2} \tilde{T}_{\mu\nu},
\end{eqnarray}
with
\begin{eqnarray}
\tilde{T}^{\mu}_{\nu}=diag(\tilde{\rho},-\tilde{p}_{r},-\tilde{p}_{t},-\tilde{p}_{t}),
\end{eqnarray}
we arrive at
\begin{eqnarray}
      k^2 \tilde{\rho} &=& \frac{1}{r^2} +
        e^{-\lambda}\!\left(\frac{\lambda'}{r} - \frac{1}{r^2}\right) \!,\label{mgde05}\\
          k^2 \tilde{p_{r}} &=& -\frac{1}{r^2} +
        e^{-\lambda}\!\left(\frac{\nu'}{r} + \frac{1}{r^2}\right)\!,\label{mgde06}\\
          k^2 \tilde{p_{t}} &=& \frac{e^{-\lambda}}{4}\!\left(2\nu'' + {\nu'}^2 - \lambda'\nu' + 2\frac{\nu'-\lambda'}{r} \right)\!\label{mgde07}.
\end{eqnarray}

Now, let us assume a splitting of the matter sector
\begin{equation}\label{energy-momentum}
    \tilde{T}_{\mu\nu} = T_{\mu\nu} + \theta_{\mu\nu}\;,
\end{equation}
with
\begin{eqnarray}
T^{\mu}_{\nu}&=&\textnormal{diag}(\rho,-p_{r},-p_{t},-p_{t})\label{tmunu}\\
\theta^{\mu}_{\nu}&=&\textnormal{diag}(\theta_{0}^{0},\theta_{1}^{1},\theta_{2}^{2},\theta_{2}^{2}),\label{thetamunu}
\end{eqnarray}
and a metric deformation
\begin{eqnarray}
    \nu\;\; &\longrightarrow &\;\; \xi +  g,\label{comp-temporal} \\
    e^{-\lambda}\;\; &\longrightarrow &\;\; e^{-\mu} +  f\;,\label{comp-radial}
\end{eqnarray}
where $\{f,g\}$ are the so-called decoupling functions. Replacing (\ref{energy-momentum}), (\ref{comp-temporal}) and (\ref{comp-radial}) in (\ref{mgde05}), (\ref{mgde06}) and (\ref{mgde07}), we arrive at
\begin{eqnarray}
k^2 \rho &=& \frac{1}{r^2}-e^{-\mu}\left(\frac{1}{r^2}-\frac{\mu'}{r}\right),\label{rhos} \\
k^2 p_r &=& -\frac{1}{r^2}+e^{-\mu}\left(\frac{1}{r^2}+\frac{\xi'}{r}\right),\label{prs}\\
k^2 p_{t} &=& \frac{e^{-\mu}}{4}\left(2\xi''+\xi'^2-\mu'\xi'+2\frac{\xi'-\mu'}{r}\right),\label{pts}
\end{eqnarray}
and 
\begin{eqnarray}
k^2\theta_{0}^{0} &=& -\frac{ f}{r^2}-\frac{f'}{r}\label{theta0}\\
k^2\theta_{1}^{1}+ Z_{1} &=& - f\left(\frac{1}{r^2}+\frac{\nu'}{r}\right)\label{theta1}\\
k^2\theta_{2}^{2}+ Z_{2} &=& -\frac{ f}{4}\left(2\nu''+\nu'^2+\frac{2\nu'}{r}\right)\nonumber\\
&& -\frac{ f'}{4}\left(\nu'+\frac{2}{r}\right),\label{theta2}
\end{eqnarray}
where 
\begin{eqnarray}
Z_{1}&=&\frac{e^{-\mu}g'}{r}\\
4Z_2&=&e^{-\mu}(2g''+ g'^2+\frac{2g'}{r}+2\xi'g'-\mu'g').
\end{eqnarray}
At this point some comments are in order. First, Eqs. (\ref{rhos}), (\ref{prs}) and (\ref{pts}) are automatically satisfied whenever $\{\xi,\mu\}$ are supplied as a seed of the system. In this regard, Eqs. (\ref{theta0}), (\ref{theta1})
 and (\ref{theta2}) must be solved for $\{f,g,\theta^{0}_{0},\theta^{1}_{1},\theta^{2}_{2}\}$ after providing two auxiliary conditions to close the system of differential equations. 
 Two conditions that have been broadly used in the construction of interior solutions by GD are the mimic constraint for the pressure, namely $p_{r}=-\beta \theta^{1}_{1}$, and the mimic constraint for the density, $\rho=\beta \theta^{0}_{0}$, with $\beta$ a constant. In general, each condition must be supplemented with an extra constraint in order to close the system. Note that, the mimic constraint of the density only involve the radial deformation $f$ so that, finding analytical solutions in this case will depend on the choice of the seed. In contrast, the mimic constrain for the pressure involve both deformation functions $\{f,g\}$ and, in most of the cases, it is not straightforward to obtain solutions of the coupled system of differential equations. As we shall see in the next section, our protocol demands for equations of state which only involve $f$. 
 
 Second, it is worth noticing that the covariant conservation of the total energy momentum tensor leads to
\begin{eqnarray}
\label{con22}
\nabla_\sigma\theta^{\sigma}_{\ \nu}
=
\frac{g'}{2}\left(\rho+p_{r}\right)\delta^{\sigma}_{\ \nu}
\ .
\end{eqnarray}
which encodes the information of energy-momentum exchange $\Delta\,{\rm E}$  between the sources, namely 
\begin{eqnarray}
\label{exchange}
\Delta\,{\rm E}=
\frac{g'}{2}\left(\rho+p_r\right)
\ ,
\end{eqnarray}
which we can write in terms of pure geometric functions as
\begin{eqnarray}
\label{exchange2}
\Delta\,{\rm E}=
\frac{g'}{2\,k^2}\frac{e^{-\mu}}{r}\left(\xi'+\mu'\right)
\ .
\end{eqnarray}
From the expression~\eqref{exchange} we can see that $g'>0$ yields $\Delta\,{\rm E}>0$. This indicates $\nabla_\sigma\theta^{\sigma}_{\ \nu}>0$,  according to the conservation equation~\eqref{con22}, which means that the source $\theta_{\mu\nu}$ is giving energy to the environment. The opposite occurs when $g'<0$.

Finally, in order to take into account that the metric should be continuous at surface $\Sigma$ of the star we have to match smoothly the interior metric with the outside, namely
\begin{eqnarray}
    e^{\nu}\Big|_{\Sigma^{-}} &=& \left(1 - \frac{2M}{r}\right)\Bigg|_{\Sigma^{+}}\,,\label{mgde11a}\\
    e^{\lambda}\Big|_{\Sigma^{-}} &=& \left(1 - \frac{2M}{r}\right)^{-1}\Bigg|_{\Sigma^{+}}\,,\label{mgde11b}\\
     \tilde{p}_r(r) \Big|_{\Sigma^{-}} &=& \tilde{p}_r(r) \Big|_{\Sigma^{+}}\,,\label{mgde11c}
\end{eqnarray}
which corresponds to the continuity of the first and second fundamental form across that surface. \\

In the next section we introduce the complexity factor and establishes a protocol to construct anisotropic solutions with vanishing complexity thorough GD.

\section{Complexity factor and the vanishing complexity condition}\label{CF}
\label{sec3}
The complexity factor introduced in \cite{Herrera:2018bww}
has been defined in the context of general relativity. This quantity is a well established concept which, for static and self gravitating spheres reads
\begin{eqnarray}\label{YTF2}
Y_{TF} = 8\pi \Pi - \frac{4\pi}{r^3}\int^{r}_{0} \tilde{r}^3 \tilde{\rho}' d\tilde{r},
\end{eqnarray}
with $\Pi=\tilde{p}_{r}-\tilde{p}_{t}$, and 
captures the essence of the least complex system corresponding to that supported by a homogeneous and isotropic fluid. In this case, $Y_{TF}=0$ and corresponds to a system with vanishing complexity. However, it is worth noticing that the complexity factor $Y_{TF}$, also vanish for all configurations where 
\begin{eqnarray} \label{YTF=0}
\Pi = \frac{1}{2 r^3}\int^{r}_{0} \tilde{r}^3 \tilde{\rho}' d\tilde{r}.
\end{eqnarray}
which may be regarded as a non–local equation of state, so we can use it to impose a plausible condition on the physical variables when solving the Einstein equations. \\

Note that by using Einstein's equation, Eq. (\ref{YTF2}) can be written as
\begin{eqnarray}\label{ytfcal}
Y_{TF}=\frac{e^{-\lambda } \left(\nu ' \left(r \lambda '-r \nu '+2\right)-2 r \nu ''\right)}{4 r},
\end{eqnarray}
so that $Y_{TF}=0$ implies
\begin{eqnarray}
\nu ' \left(r \lambda '-r \nu '+2\right)-2 r \nu ''=0,
\end{eqnarray}
which can be expressed as a total derivative
\begin{eqnarray}
\frac{d}{dr}\left(\log\nu'+\frac{\nu-\lambda}{2}-\log r\right)=0.
\end{eqnarray}
Next, we arrive at
\begin{eqnarray}
\log\nu'+\frac{\nu-\lambda}{2}-\log r=constant,
\end{eqnarray}
from where, taking the exponential function at both sides leads to
\begin{eqnarray}
e^{\nu/2}\nu'=\tilde{\alpha} r e^{\lambda/2}
\end{eqnarray}
with $\tilde{\alpha}$ a constant. Now, if we define,
\begin{eqnarray}\label{nuyu}
\nu=2\log u
\end{eqnarray}
we obtain
\begin{eqnarray}\label{u}
u'=\alpha r e^{\lambda/2},
\end{eqnarray}
with $\alpha=\tilde{\alpha}/2$. In this regard, after solving Eq. (\ref{u}), we can obtain $\nu$ analytically from Eq. (\ref{nuyu}). This procedure is advantageous in the context MGD-e where $\nu=\xi+g$, so that
\begin{eqnarray}\label{gyu}
g=2\log u-\xi.
\end{eqnarray}
At this point a couple of comments are in order. First, note that Eq. (\ref{gyu}) allows to obtain $g$ in terms of the seed $(\xi,\mu)$ and the radial deformation $f$. Of course, this requires the previous computation of the radial deformation by assuming some equation of state which only involve $f$ and its derivatives. Now, that Eq. (\ref{gyu}) admits analytical solutions or not will depend critically on the choice of both the seed and the equation of state. Second, once the integration is accomplished, the total solution automatically fulfills the vanishing complexity condition by construction.\\
Based on the above results, let us propose the following protocol to obtain interior solutions with vanishing complexity by GD through MGD-e:
\begin{enumerate}
\item Consider a seed metric $\{\xi,\mu\}$.\\

\item Impose an equation of state which only involve $f(r)$, in particular, the mimic constrain for the density (for example).\\

\item With $f$, construct $e^{-\lambda}=e^{-\mu}+f$ and replace it in Eq. (\ref{u}) to solve for $u(r)$. \\

\item Use the result obtained for $u(r)$ to obtain $\nu$ from (\ref{nuyu}).
\end{enumerate}
In the next sections, we shall implement the protocol using the Tolman VII and Tolman IV solution as seeds.

\section{Like--Tolman VII solution}\label{tolman}
Recently, the Tolman VII solution \cite{Tolman:1939jz} has grabbed the attention in the community given that it allows to represent neutron stars with realistic equations of state \cite{Neary:2001ai,Jiang:2019vmf,Jiang:2020uvb,Hod:2018mvu,Hladik:2020xfw,Posada:2020svn}, contrary to the famous internal Schwarzschild solution which is sourced by an uniform distribution of energy \cite{Schwarzschild:1916ae,Stuchlik:2000gey}. An anisotropic version of the Tolman VII solution was reported in \cite{Hensh:2019rtb}. Besides, 
extremely compact Tolman VII solutions were treated in \cite{Posada:2020svn} while the exact version was recently found by numerical solution of the Einstein equations in \cite{Posada:2022lij}.

In this section we shall use the Tolman VII solution as a seed
$\{\xi,\mu,\rho, p\}$ for perfect fluids~\cite{Tolman:1939jz}, namely, 
\begin{eqnarray}
\label{tolman00}
&&e^{\xi}
=
B^2 \sin ^2\log \left(\sqrt{\frac{\exp \left(-\frac{\mu (r)}{2}\right)+\frac{2 r^2}{A^2}-\frac{A^2}{4 R^2}}{C}}\right)
\\
\label{tolman11}
&&e^{-\mu}
=
1-\frac{r^2}{R^2}+\frac{4 r^4}{A^4} 
\\
\label{tolmandensity}
&&\rho
=
\frac{1}{8 \pi }\left(\frac{3}{R^2}-\frac{20 r^2}{A^4}\right),
\\
&&p=\frac{
4 A^2 R^2 \Xi(r) \cot\Psi(r) -A^4+4 r^2 R^2}{8 \pi  A^4 R^2}
\label{tolmanpressure}\\
&&\Xi=\sqrt{\frac{4 r^4}{A^4}-\frac{r^2}{R^2}+1}\\
&&\Psi=\frac{1}{2} \log \left(\frac{\Xi(r)+\frac{2 r^2}{A^2}-\frac{A^2}{4 R^2}}{C}\right),
\end{eqnarray}
where $R$ is the radius of the stellar configuration and
the constants $A$, $B$ and $C$ are determined by the matching conditions in Eqs.~(\ref{mgde11a}), (\ref{mgde11b}) 
and~(\ref{mgde11c}) with $f_R=g_R=0$, which leads to
\begin{eqnarray}
\label{A}
A&=&\left(\frac{4R^{5}}{R-2M}\right)^{1/4}\\
B&=&\sqrt{\frac{5M^{2}-4MR+R^{2}}{R(R-2M)}}\\
C&=&\frac{(3 R-8 M) e^{-2 \cot ^{-1}\left(\frac{M}{R-2 M}\right)}}{2  \sqrt{R(R-2 M)}},
\end{eqnarray}
with the compactness $M/R<4/9$, and $M$ the total mass.
The expressions in Eq.~(\ref{A}) ensure the geometric continuity at $r=R$. 

Following the protocol in section \ref{CF}, the next step is to consider an equation of state which only involve $f(r)$. We note that, given the simple form of (\ref{tolmandensity}), in this case is appropriate to implement the mimic constraint for the density, namely
\begin{eqnarray}\label{mcd}
\theta^{0}_{0}=\beta\rho,
\end{eqnarray}
with $\beta$ a constant, from where
\begin{eqnarray}\label{diffb}
\frac{f'}{r}+\frac{f}{r^2}=\beta\left(\frac{20r^2}{A^4}-\frac{3}{R^2}\right).
\end{eqnarray}
Solving (\ref{diffb}) leads to
\begin{eqnarray}
f=\beta  \left(\frac{4 r^4}{A^4}-\frac{r^2}{R^2}\right),
\end{eqnarray}
so that the total radial metric reads
\begin{eqnarray}
e^{-\lambda}=1+(1+\beta)  \left(\frac{4 r^4}{A^4}-\frac{r^2}{R^2}\right).
\end{eqnarray}
Replacing in (\ref{u}), we obtain
\begin{eqnarray}
u=\frac{\alpha  A^2}{2 \left(4 \sqrt{\beta +1}\right)} \log \left(
\frac{\eta_{+}+A^4}{\eta_{-}-A^4}\right)+c_{1}
\end{eqnarray}
with
\begin{eqnarray}
\eta_{\pm}=4 R \left(\sqrt{A^4 \left(\frac{R^2}{\beta +1}-r^2\right)+4 r^4 R^2}\pm2 r^2 R\right),
\end{eqnarray}
from where $\nu$ and $g$ can be easily determined through (\ref{nuyu}) and (\ref{gyu}). After implementing the matching conditions, all the parameters can be expressed in terms of $\{R,M,\beta\}$, namely
\begin{eqnarray}
A&=&\frac{\sqrt{2} \sqrt[4]{\beta +1} R^{5/4}}{\sqrt[4]{-2 M+\beta  R+R}}\\
c_{1}&=&-\frac{M \log \left(\frac{\Theta _+}{\Theta _-}\right) \sqrt{\frac{R^6 (R-2 M)}{-2 M+\beta  R+R}}+4 R^3 (R-2 M)}{4 R^{7/2} \sqrt{R-2 M}}\\
\alpha &=&\frac{4 \left(c_1 \sqrt{R}+\sqrt{R-2 M}\right) \sqrt{-2 M+\beta  R+R}}{R^3 \log \left(\frac{\Theta _+}{\Theta _-}\right)}
\end{eqnarray}
with
\begin{eqnarray}
\Theta_{\pm}&=&4 M \left(\sqrt{\frac{R^6 (R-2 M)}{-2 M+\beta  R+R}}-R^3\right)\nonumber\\
&&\pm(\beta +1) R \left(R^3-2 \sqrt{\frac{R^6 (R-2 M)}{-2 M+\beta  R+R}}\right)
\end{eqnarray}

\begin{figure*}[ht!]
	\centering
	\includegraphics[width=0.45\textwidth]{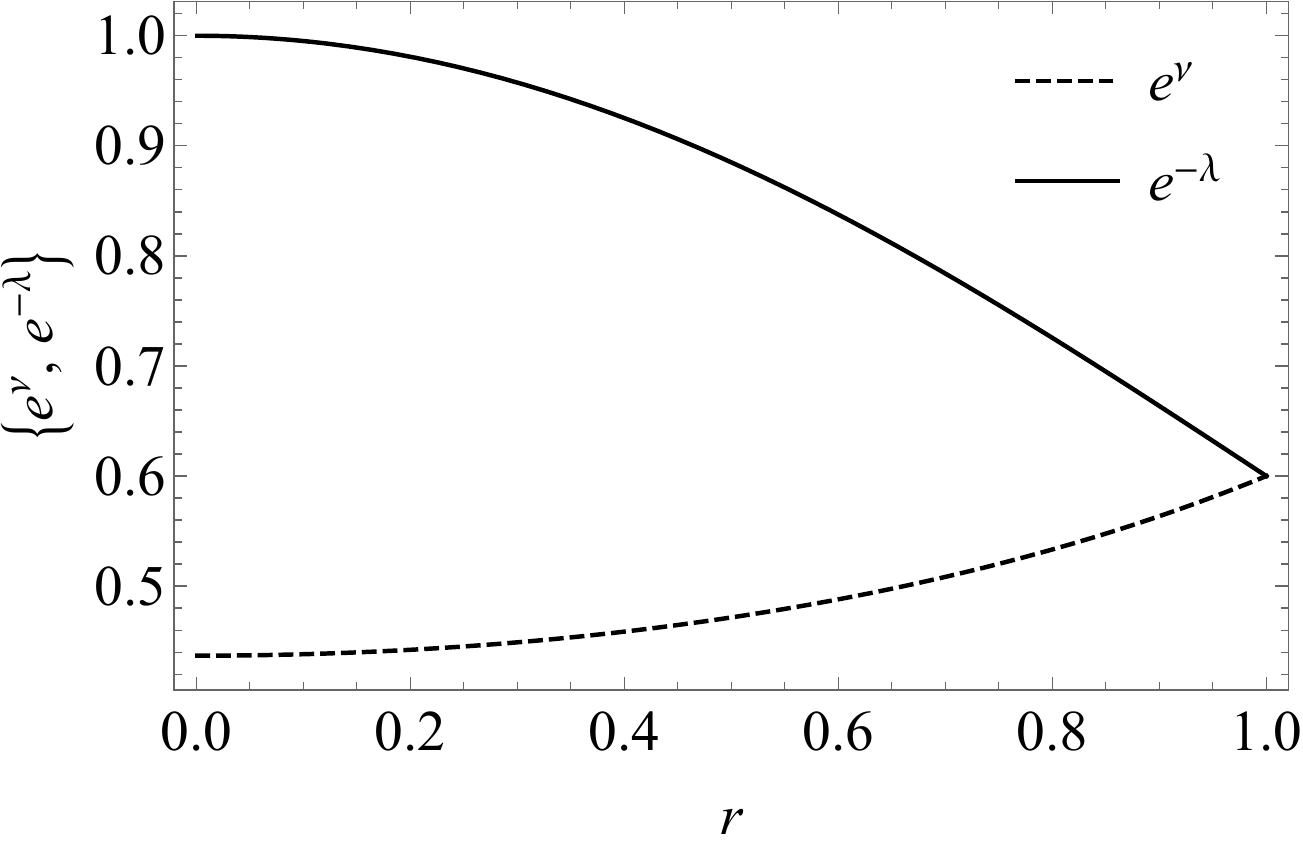} \
	\includegraphics[width=0.45
	\textwidth]{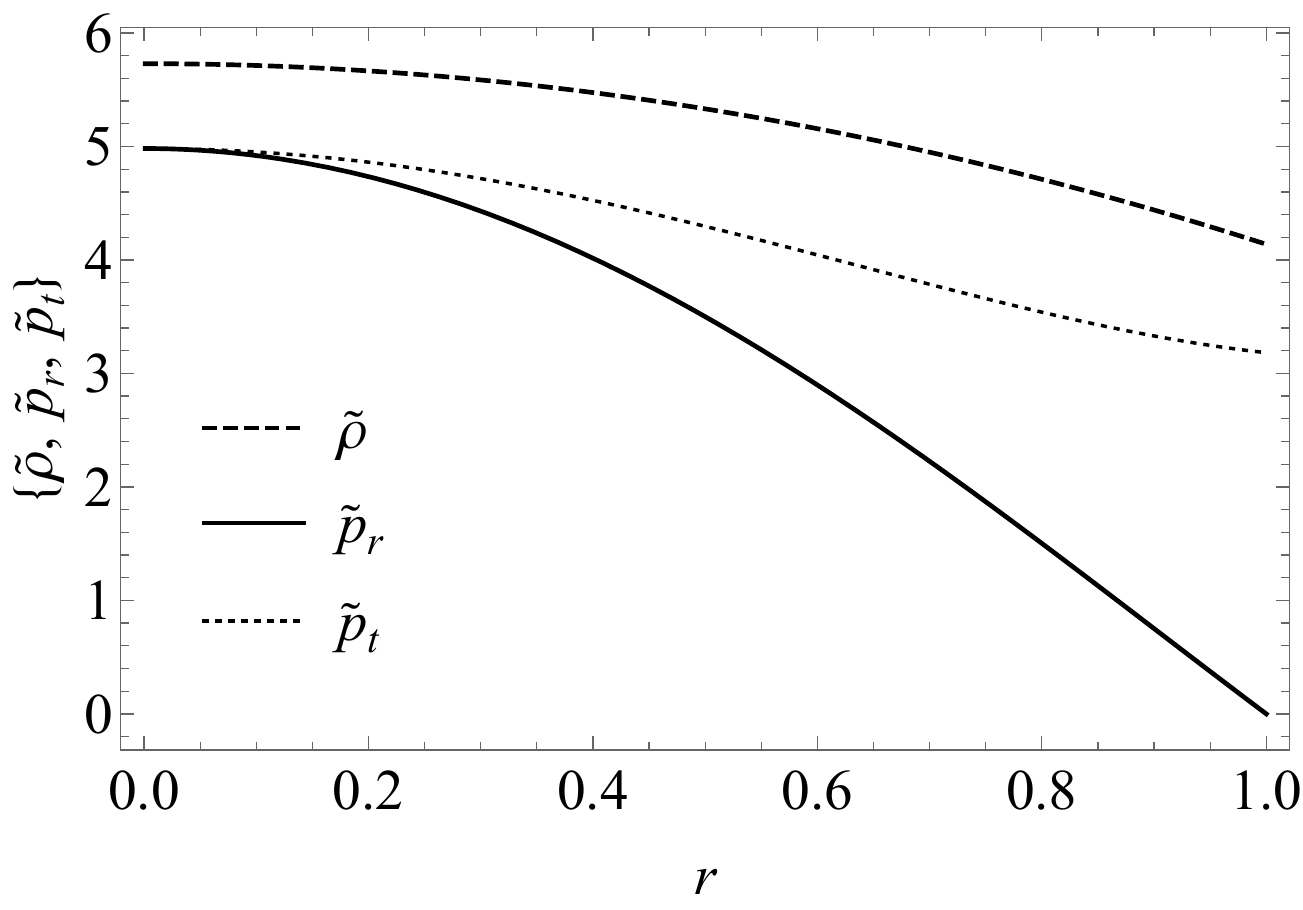} \
	\caption{Metric functions $\{e^{\nu},e^{-\lambda}\}$ (left panel) and matter sector $\{10^{2}\times\tilde{\rho}$,$10^{3}\times\tilde{p}_{r},10^{3}\times\tilde{p}_{t}\}$ (right panel) as a function of the radial coordinate $r$, for $M/R=0.2$ and $\beta=-0.52$ for the like--Tolman VII model.}
	\label{fig1}.
\end{figure*}

The obtained solution satisfies the basic requirements of an interior solution as shown in figure \ref{fig1}.

Another interesting point that can be discussed is the exchange of energy between the generic fluid $\theta_{\mu\nu}$ and the perfect fluid given by 
Eq. (\ref{exchange2}). First of all, we note that although the total solution does not depends on the parameter $C$ appearing in the seed metric (\ref{tolman00}), the expression for the exchange of energy does so that such a parameter is free. The interesting feature is that, depending on its values, we observe a transition between situations in which $\Delta E>0$ (generic fluid giving energy) to situations where $\Delta E<0$ (perfect fluid giving energy) as shown in figure \label{energy}
\begin{figure*}[ht!]
	\centering
	\includegraphics[width=0.45\textwidth]{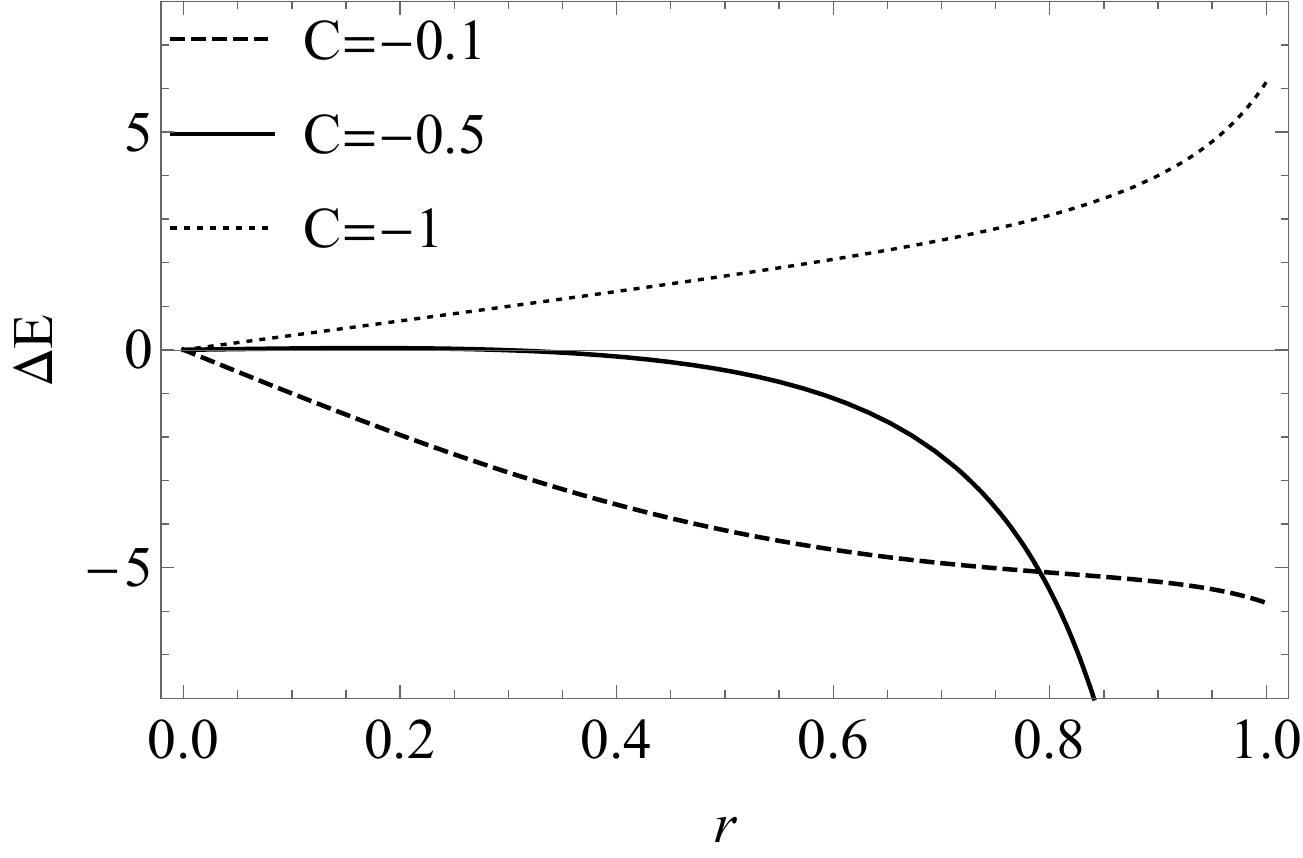} \
	\hspace{1cm}
	\includegraphics[width=0.35
	\textwidth]{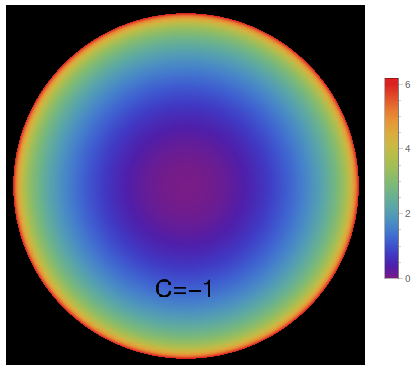} \
	\caption{Left panel: Exchange of energy $10^{2}\times\Delta E$ 
for different values of $C$. Right panel: Density plot of the exchange of energy	for $C=-1$ for the like--Tolman VII model. In both cases we have taken $M/R=0.2$ and $\beta=-0.52$ .}
	\label{energy}.
\end{figure*}%

\section{Like--Tolman IV solution}\label{tolman4}
In this section we shall consider the Tolman IV solution \cite{Tolman:1939jz} as a seed to test our method with a different condition involving $f$. In this case the metric reads
\begin{eqnarray}
\label{tolman001}
&&e^{\xi}
=
B^{2}\left(1+\frac{r^{2}}{A^{2}}\right)
\label{tolman111}\\
&&e^{-\mu}
=\frac{(C^{2}-r^{2})(A^{2}+r^{2})}{C^{2}(A^{2}+2r^{2})}
\\
\label{tolmandensity1}
&&\rho
=
\frac{3A^{4}+A^{2}(3C^{2}+7r^{2})+2r^{2}(C^{2}+3r^{2})}{8 \pi C^{2}(A^{2}+2r^{2})^{2} },
\\
&&p=\frac{C^{2}-A^{2}-3r^{2}}{8 \pi  C^{2}(A^{2}+2r^{2})}
\label{tolmanpressure1}\\
\end{eqnarray}
where $R$ is the radius of the stellar configuration and
the constants $A$, $B$ and $C$ are determined by the matching conditions in Eqs.~(\ref{mgde11a}), (\ref{mgde11b}) 
and~(\ref{mgde11c}) with $f_R=g_R=0$

By simple inspection of Eq. (\ref{tolmandensity1}), we note that the imposition of the mimic constraint for the density is not convenient in this case. Instead, we shall impose the mimic constraint for the mass function \footnote{Indeed, the mimic constraint for the mass in the volume integral of the mimic constraint for the density. They are equivalent, but the implementation of the former is straightforward because we avoid to solve a differential equation which should lead to the same result}, namely 
\begin{eqnarray}
m(r)=\beta m_{\theta}(r),
\end{eqnarray}
which (taking $\beta=1$ for simplicity) leads to
\begin{eqnarray}\label{mcd1}
\frac{r}{2}(1-e^{-\mu})=-\frac{r}{2}f
\end{eqnarray}
Now, replacing (\ref{tolman111}) in (\ref{mcd1}) we obtain
\begin{eqnarray}
f=-\frac{r^2 \left(A^2+c^2+r^2\right)}{c^2 \left(A^2+2 r^2\right)}.
\end{eqnarray}
It is worth mentioning that, as far as we known, this is the first time that such a constraint is considered. Besides, it results advantageous in the sense that the radial deformation can be obtained algebraically.

Next, by solving Eq. (\ref{u}), we obtain
\begin{eqnarray}
u=\eta\ _2F_1\left(1,\frac{5}{4};\frac{7}{4};\frac{\left(A^2+2 r^2\right)^2}{A^4+2 c^2 A^2}\right)+c_{1},
\end{eqnarray}
with $_2F_1\left(a,b;c;z\right)$ the hypergeometric function and
\begin{eqnarray}
\eta=\frac{\alpha  C^2 \left(A^2+2 r^2\right)^2 \sqrt{\frac{A^2 \left(c^2-2 r^2\right)-2 r^4}{c^2 \left(A^2+2 r^2\right)}} \, }{3 A^2 \left(A^2+2 c^2\right)}.
\end{eqnarray}
Following the protocol, $\nu$ and $g$ can be easily determined through (\ref{nuyu}) and (\ref{gyu}). After implementing the matching conditions, all the parameters can be expressed in terms of $\{R,M,C\}$ but the expression are too long to be shown here.

\begin{figure*}[ht!]
	\centering
	\includegraphics[width=0.45\textwidth]{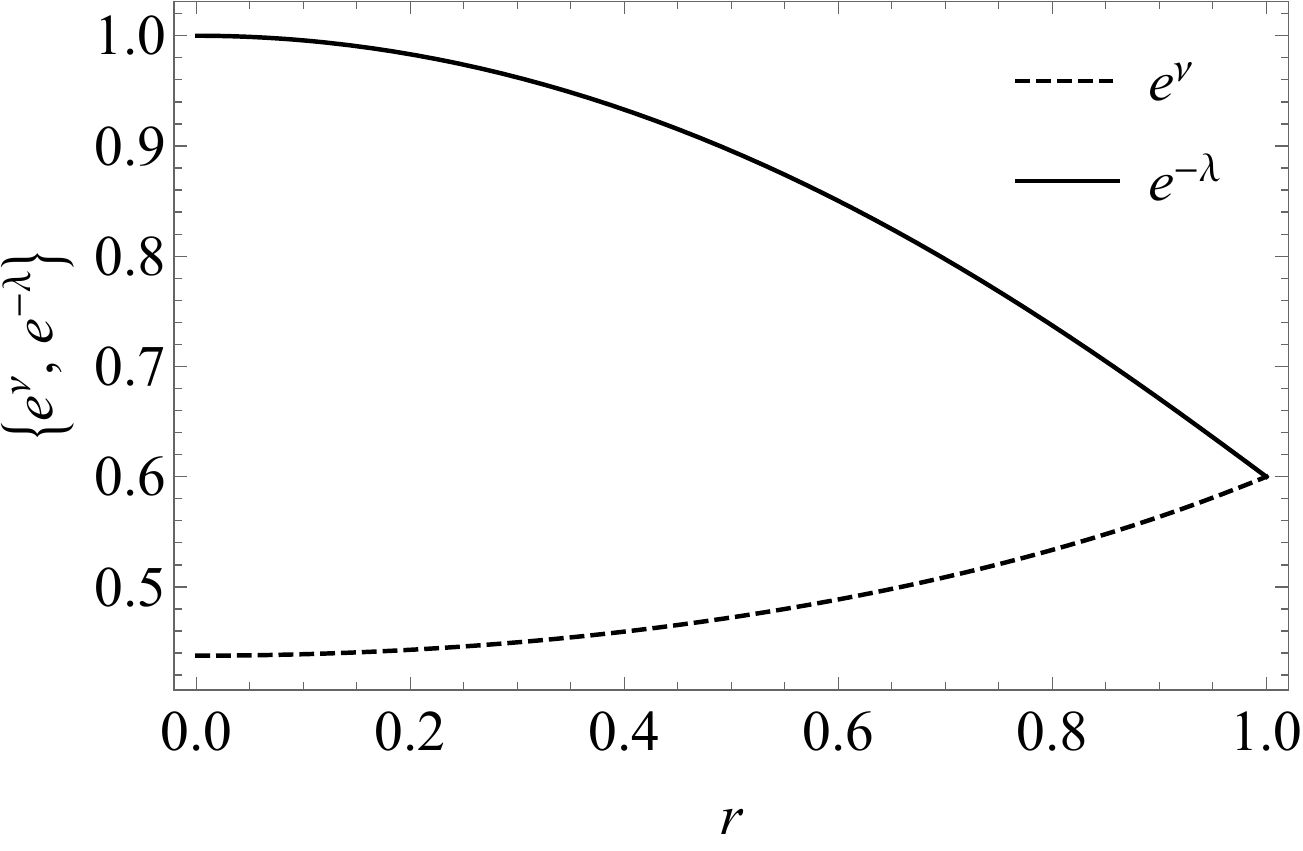} \
	\includegraphics[width=0.45
	\textwidth]{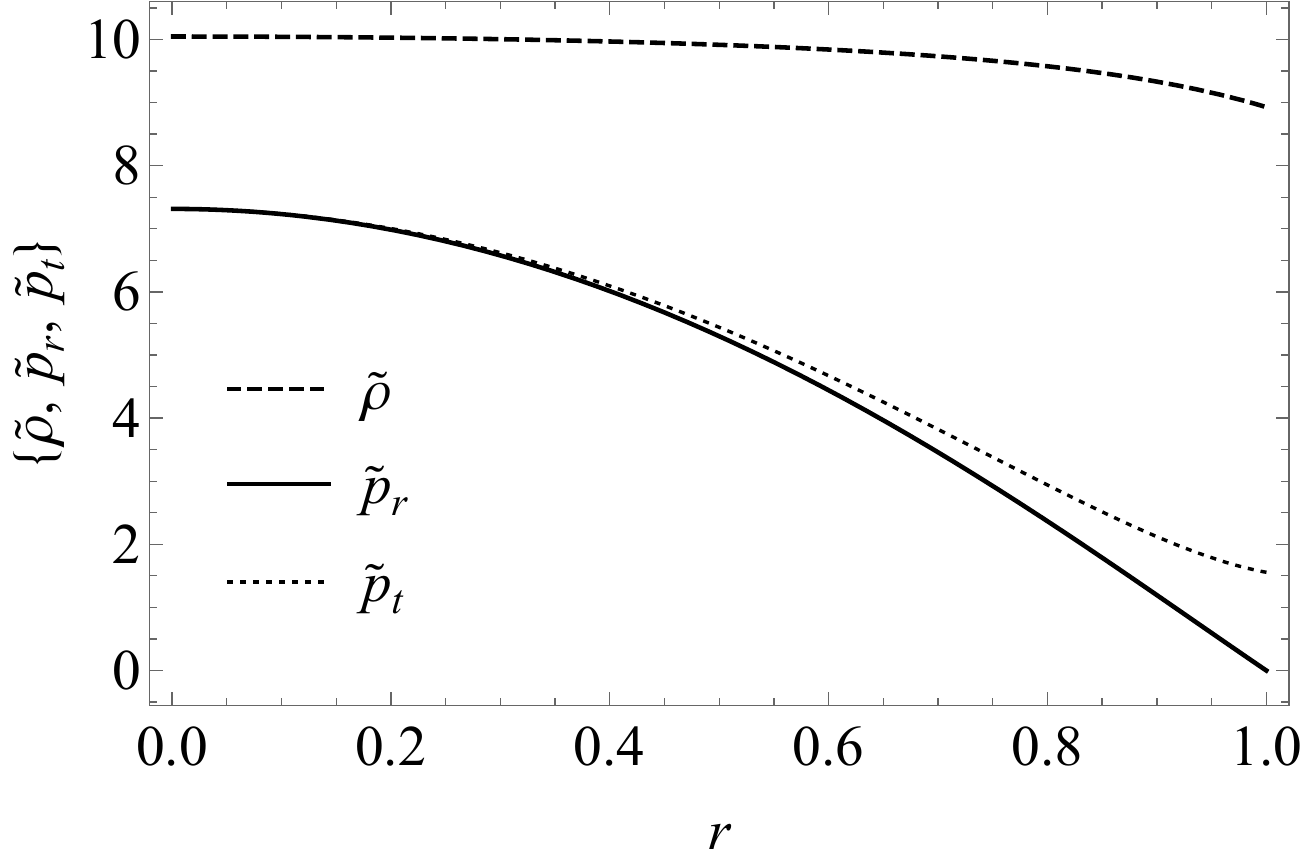} \
	\caption{Metric functions $\{e^{\nu},e^{-\lambda}\}$ (left panel) and matter sector $\{10^{2}\times2\tilde{\rho}$,$10^{3}\times\tilde{p}_{r},10^{3}\times\tilde{p}_{t}\}$ (right panel) as a function of the radial coordinate $r$, for $M/R=0.2$ and $C=1.5$ for the like--Tolman IV model.}
	\label{fig11}.
\end{figure*}

It is worth noticing that the solution satisfies the basic requirements of an interior solution as shown in figure \ref{fig11} whenever $C\approx 1.5$. \footnote{A careful numerical analysis not shown here reveals that $C\in(1.3,1.55)$ }

\begin{figure*}[ht!]
	\centering
	\includegraphics[width=0.45\textwidth]{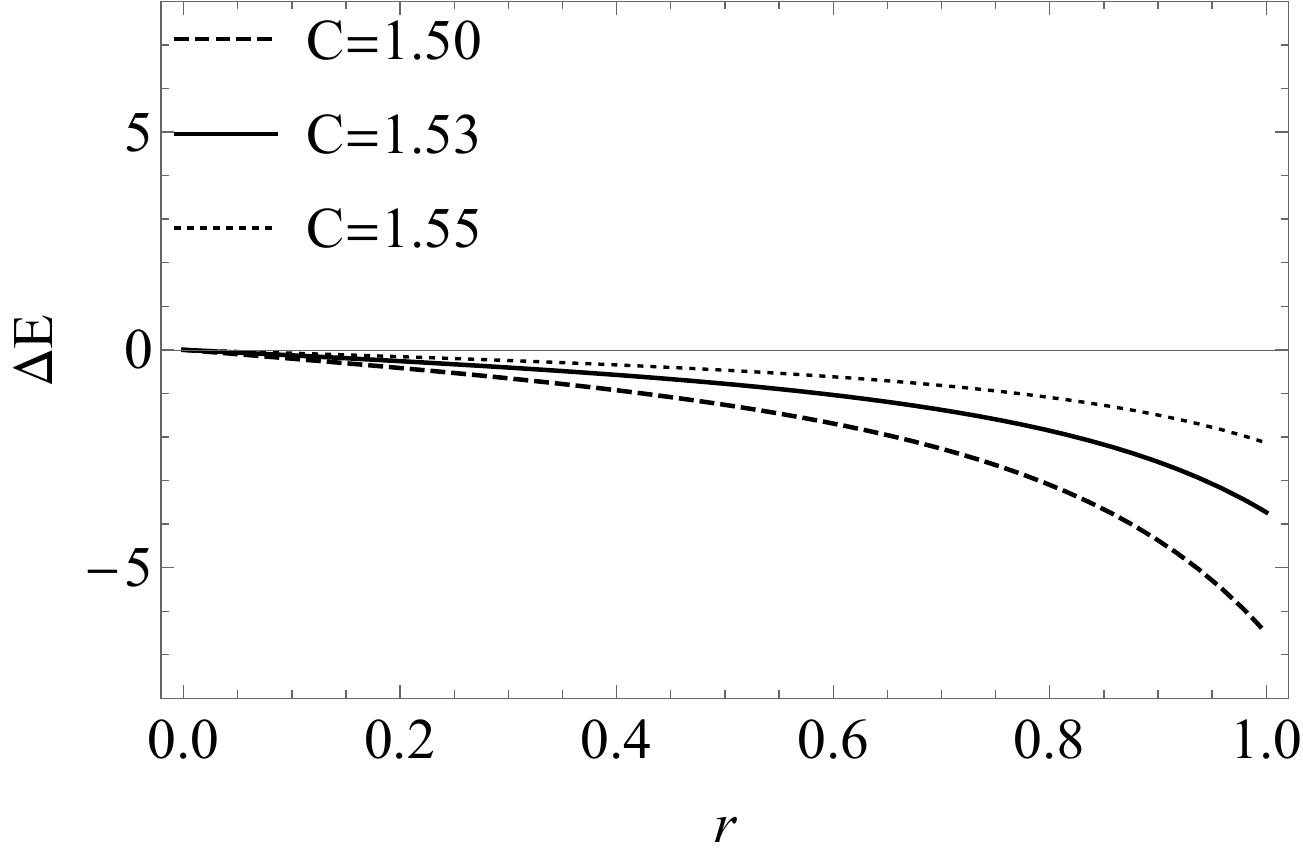} \
	\hspace{1cm}
	\includegraphics[width=0.35
	\textwidth]{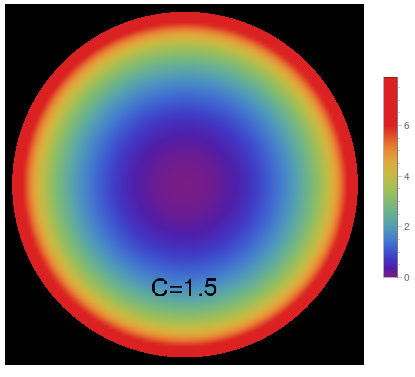} \
	\caption{Left panel: Exchange of energy $10^{3}\times2\Delta E$ 
for different values of $C$. Right panel: Density plot of the exchange of energy	for $C=1.5$ for the like--Tolman IV model. In both cases $M/R=0.2$.}
	\label{e-energy2}.
\end{figure*}

The exchange of energy is shown in Fig. \ref{e-energy2}. In contrast to the previous case, in this case the physical solution has associated $\Delta E<0$ which corresponds to the situation in which the generic source gives energy to to the perfect fluid.

\section{Conclusions}
The Gravitational Decoupling by the Minimal Geometric Deformation is a invaluable tool in the construction of well--behaved solutions. However, it extended version, requires dealing with non--trivial differential equations for the deformation metrics and sometimes obtaining analytical solutions for both functions is simply not possible. In this regard, in this work we proposed a simple protocol to simplify the differential equations arising from the Gravitational Decoupling by the Minimal Geometric Deformation extended. More precisely, we used the well--established definition of complexity for self--gravitating spheres to obtain a simple constraint between the radial and the temporal components of a general metric for static and spherically symmetric spacetimes. It is worth mentioning that our method works whenever the supplementary condition only involve the radial deformation.

In order to test our protocol, we used both the Tolman VII and Tolman IV solutions as seeds.  In the former case, we used the mimic constraint for the density which (as required by the method) only involve the deformation $f$. In this case, given the simple form of the radial metric, we obtained analytical expressions for the total solution. However, in latter case, the mimic constraint is not suitable for the implementation of the protocol so we used a mimic constraint for the mass function that, as far as we know, has not been explored before. We found that this constraint leads to (semi) analytical expressions for the total solution involving hypergeometric functions. In both cases, we  obtain new anisitropic systems with nonvaishing complexity satisfying the basic requirements of  an interior solution. Additionally, we obtained the exchange of energy between the perfect fluid and the generic source $\theta_{\mu\nu}$ and, in contrast to the results reported in previous works, in our case the generic fluid can give or receive energy from the perfect fluid in the case of Tolman VII depending on the value taken by the free parameter involved. Besides, the generic fluid only give energy to the perfect fluid in the case of Tolman IV. 

\subsection*{Acknowledgments}
EC is suported by Polygrant ${\rm N}^{\rm o}$  17459

%
%\subsection*{Acknowledgments}
%

%\section*{References}
\bibliography{arxiv-file.bib}
\bibliographystyle{unsrt}

%%%%%%%%%%%%%%%%%%%%%%%%%%%%%%%%%%%%%%%%%%%%%%%%%%%%%%%%%
%\section{Acknowledgements}--------------------------------------------------------
\end{document}